\documentclass[10pt,twocolumn,letterpaper]{article}

\usepackage[pagenumbers]{cvpr} 

\usepackage{graphicx}
\usepackage{amsmath}
\usepackage{amssymb}
\usepackage{booktabs}
\usepackage{color}

\usepackage[pagebackref,breaklinks,colorlinks,citecolor=magenta]{hyperref}

\usepackage[capitalize]{cleveref}
\crefname{section}{Sec.}{Secs.}
\Crefname{section}{Section}{Sections}
\Crefname{table}{Table}{Tables}
\crefname{table}{Tab.}{Tabs.}

\def\cvprPaperID{38} 
\def\confName{CVPR}
\def\confYear{2023}

\begin{document}
	
	\title{Large Kernel Distillation Network for Efficient Single Image Super-Resolution}
	
	\author{Chengxing Xie\textsuperscript{1}\footnotemark[1], Xiaoming Zhang\textsuperscript{1}\footnotemark[1], Linze Li\textsuperscript{1}, Haiteng Meng\textsuperscript{1}, Tianlin Zhang\textsuperscript{2}, Tianrui Li\textsuperscript{1}, Xiaole Zhao\textsuperscript{1}\footnotemark[2]\\
		\textsuperscript{1}Southwest Jiaotong University, China\\
		\textsuperscript{2}National Space Science Center, Chinese Academy of Science, China\\
		{\tt\small zxc0074869@gmail.com, zxiaoming360@gmail.com, zxlation@foxmail.com}}
\maketitle

\renewcommand{\thefootnote}{\fnsymbol{footnote}}
\footnotetext[1]{Equal contributions to this work. $\dag$Corresponding author.}

\begin{abstract}
	Efficient and lightweight single-image super-resolution (SISR) has achieved remarkable performance in recent years. One effective approach is the use of large kernel designs, which have been shown to improve the performance of SISR models while reducing their computational requirements. However, current state-of-the-art (SOTA) models still face problems such as high computational costs. To address these issues, we propose the Large Kernel Distillation Network (LKDN) in this paper. Our approach simplifies the model structure and introduces more efficient attention modules to reduce computational cost while also improving performance. Specifically, we employ the re-parameterization technique to enhance model performance without adding extra cost. We also introduce a new optimizer from other tasks to SISR, which improves training speed and performance. Our experimental results demonstrate that LKDN outperforms existing lightweight SR methods and achieves SOTA performance. The codes are available at \href{https://github.com/stella-von/LKDN}{https://github.com/stella-von/LKDN}.
\end{abstract}

\section{Introduction}
Single image super-resolution (SISR) is an essential problem in low-level computer vision (CV) that involves reconstructing a high-resolution (HR) image from its low-resolution (LR) counterpart. After the introduction of deep learning to super-resolution by SRCNN~\cite{Dong2015Image}, there has been a significant surge in the development of deep-learning-based SR models. Due to their impressive ability to reconstruct high-resolution images from low-resolution observations, these algorithms have gained popularity in the CV community. Although deeper and larger models are often considered the optimal approach for designing SR models with strong representation ability~\cite{Lim2017Enhanced, Zhang2018Image}, there is a growing emphasis on developing lightweight models that can approximate the performance of larger models with greatly reduced parameters and less computational complexity.
\begin{figure}[!htbp]
	\centering
	\includegraphics[scale=0.24]{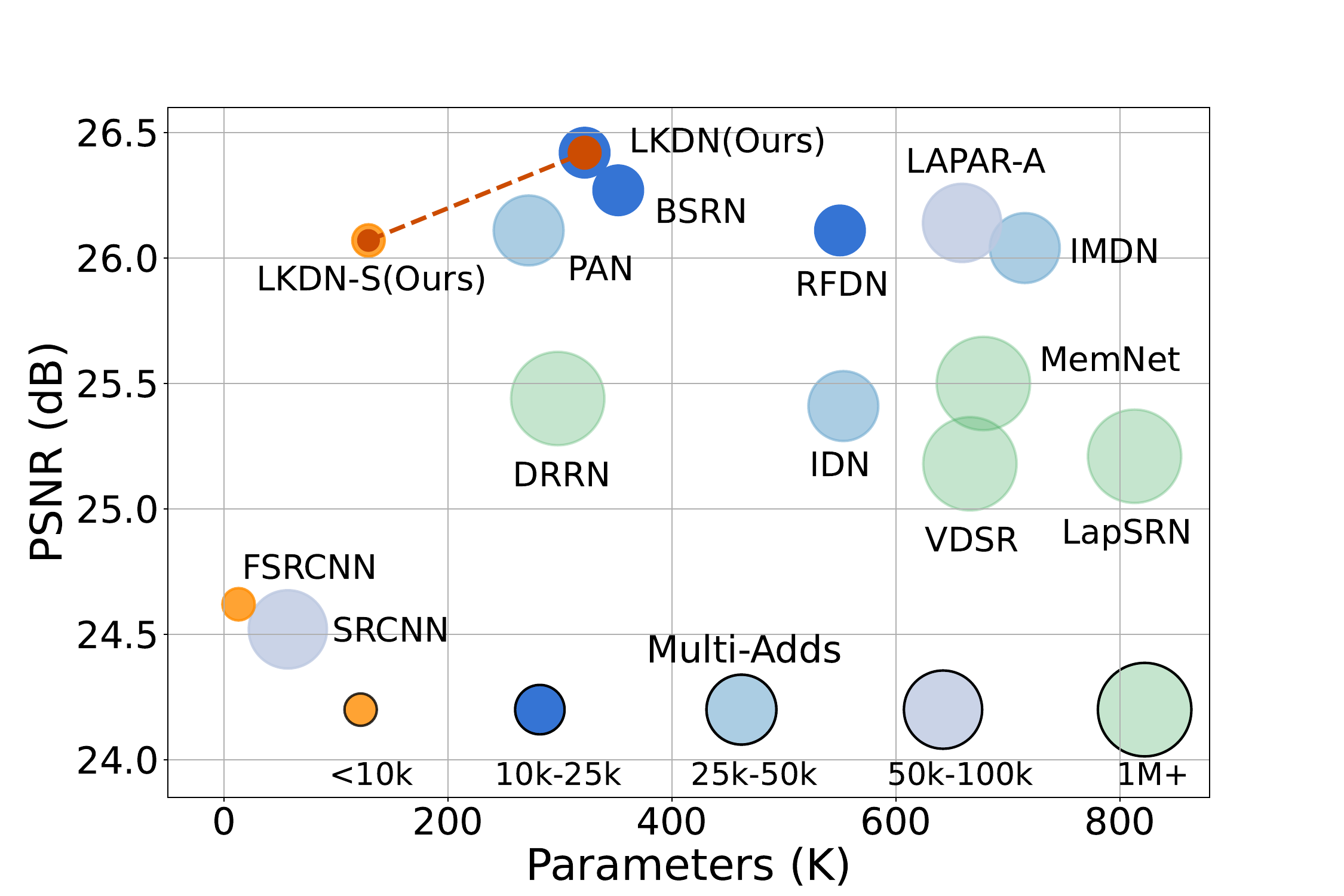}
	\caption{Comparison of model performance and complexity on Urban100~\cite{huang2015single} with SR($\times4$).}
	\label{lkdn}
\end{figure}
\vspace{-0.2cm}
Among the numerous design approaches for lightweight super-resolution models, information distillation connections~\cite{hui2019lightweight} have been identified as a highly effective method. This approach fuses features of varying hierarchies to facilitate the transmission of more distinctive features into the network, resulting in efficient feature reuse and achieving a better balance between reconstruction accuracy and computational efficiency.

RFDN~\cite{liu2020residual} reevaluated the information multi-distillation network~\cite{hui2019lightweight} and introduced multi-feature distillation connections that are highly adaptable and lightweight, and it won the champion of AIM 2020 Efficient SR Challenge~\cite{zhang2020aim}. Meanwhile, BSRN~\cite{li2022blueprint} achieved the first place in the model complexity track in NTIRE 2022 Efficient SR Challenge~\cite{li2022ntire} by incorporating residual feature distillation connections with effective attention modules and re-parameterizing the redundant convolution of RFDN by using blueprint separable convolution~\cite{haase2020rethinking} (BSConv). Besides, RFLN~\cite{Kong2022Residual} emerged as the champion of the runtime track by improving the efficiency of RFDN through the use of residual local feature blocks that reduce network fragments while maintaining model capacity. The newly proposed VAPSR~\cite{zhou2023efficient} is designed with a concise structure and fewer parameters while achieving SOTA performance. By enhancing the pixel attention mechanism~\cite{zhao2020efficient}, incorporating large kernel convolutions, and implementing efficient depth-wise separable large kernel convolutions. Additionally, VAPSR achieves comparable performance to RFDN while utilizing only 28.18\% of its parameters and outperforms BSRN.
\begin{figure*}[!htbp]
	\centering
	\includegraphics[scale=1]{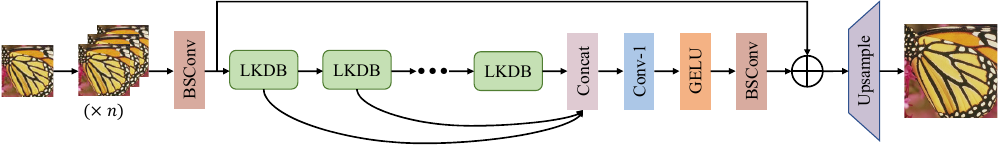}
	\caption{The architecture of large kernel distillation network (LKDN).}
	\label{lkdn}
\end{figure*}

\begin{figure*}[!htbp]
	\setlength{\abovecaptionskip}{4mm}
		\centering
	\begin{subfigure}{1\linewidth}
		\centering
		\includegraphics[scale=0.9]{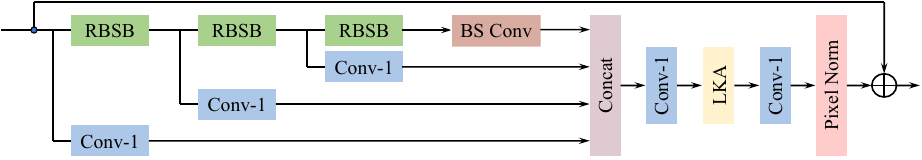}
		\caption{LKDB}
		\label{lkdb}
		\vspace{0.5cm}
	\end{subfigure}

	\begin{subfigure}{0.4\linewidth}
		\centering
		\includegraphics[scale=0.9]{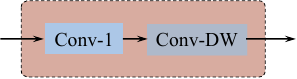}
		\caption{BSConv}
		\label{bsconv}
	\end{subfigure}
	\begin{subfigure}{0.59\linewidth}
		\centering
		\hspace{-0.8cm}
		\includegraphics[scale=0.9]{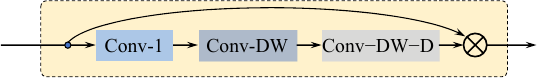}
		\caption{LKA}
		\label{lka}
	\end{subfigure}
	
	\begin{subfigure}{1\linewidth}
			\vspace{0.5cm}
		\centering
		\includegraphics[scale=0.9]{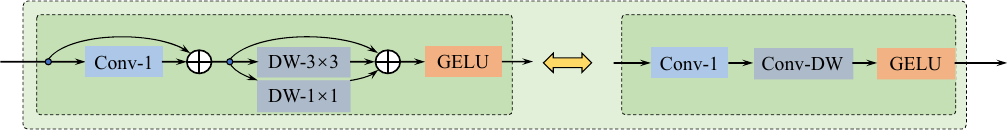}
		\caption{RBSB}
		\label{rbsb}
	\end{subfigure}
\caption{The details of each component. (a) LKDB: Large Kernel Distillation Block; (b) BSConv: Blueprint Separable Convolution; (c) LKA: Large Kernel Attention; (d) RBSB: Re-parameterized Blueprint Shallow Block.}

\end{figure*}
Although BSRN has made great progress in terms of model parameter and computation, excessive residual connections and complex attention modules (i.e. contrast-aware attention~\cite{hui2019lightweight} (CCA) and enhanced spatial attention~\cite{liu2020residual} (ESA)) have led to low model computation efficiency. As analyzed in RLFN, the complex ESA module is redundant and it is difficult to quantitatively analyze which parts are truly useful. While VAPSR achieves better performance than BSRN, the model's running speed is slower due to the presence of a large number of inefficient element-wise multiplications. By removing redundant modules in the network and introducing more efficient ones, we can build a more efficient SR network.

In this paper, we propose a novel lightweight SR network, named large kernel distillation network (LKDN), which builds upon the baseline model of BSRN. Our approach simplifies the model structure and employs a more efficient attention module, called large kernel attention (LKA), to improve model performance and computational costs. We demonstrate the effectiveness of LKA in lightweight SR tasks. Additionally, we leverage the re-parameterization technique to further enhance the performance without adding any additional computational cost. To achieve faster convergence and state-of-the-art (SOTA) performance, we incorporate the recently proposed Adan optimizer~\cite{xie2022adan}, which has shown success in various tasks such as high-level CV, NLP, and reinforcement learning. Our proposed LKDN achieves SOTA performance among existing efficiency-oriented SR networks, as shown in Figure ~\ref{lkdn}.
The main contributions of this article are:

\noindent(\textbf{1}) After analyzing the computational efficiency of BSRN~\cite{li2022blueprint} and VAPSR~\cite{zhou2023efficient}, We achieved better performance while reducing the number of parameters and computational consumption through simplifying the model structure and introducing a more efficient attention module.

\noindent(\textbf{2}) We used the technique of re-parameterization to improve the model performance without introducing any additional inference burden.

\noindent(\textbf{3}) We introduced a new optimizer that can simultaneously boost the training speed and performance of SISR models.

\section{Related Work}
\subsection{Efficient SR Models}
The development of lightweight super-resolution networks has received increasing attention in recent years due to their practical applications in resource-constrained scenarios such as mobile devices and embedded systems. Many lightweight SR models have been proposed to reduce the computational cost and memory footprint while maintaining the model capacity and achieving satisfactory performance. The common strategies for lightweight SR networks include network pruning~\cite{he2014reshaping,liu2018rethinking,zhang2021learning}, parameter sharing~\cite{Kim2016Deeply,Tai2017Image}, knowledge distillation~\cite{gao2019image,he2020fakd,zhang2021data}, depth-wise separable convolutions~\cite{sifre2014rigid,ahn2018fast, sun2022hybrid,li2022blueprint}, attention mechanisms~\cite{hui2019lightweight,liu2020residual,guo2022visual,zhao2020efficient,zhou2023efficient}, efficient upsampling methods~\cite{shi2016real,lai2017deep,tong2017image, zhao2020efficient} and re-parameterization technique~\cite{arora2018optimization, zhang2021edge}. In addition to the aforementioned techniques, in terms of network structure design, information distillation connections~\cite{hui2019lightweight,liu2020residual,Kong2022Residual,li2022blueprint} has also been demonstrated as an effective approach to building lightweight SR network architectures. We thus maintain the network topology design of information distillation while enhancing the attention mechanism and implementing re-parameterization in LKDN.

\subsection{Large Kernel Design}
Since the introduction of VGG~\cite{simonyan2014very}, small convolution kernels such as $3\times3$ have been widely used due to their high efficiency and lightweight nature. Transformer~\cite{Vaswani2017Attention}, as a model that achieves a larger receptive field through global self-attention operation, has achieved excellent performance in natural language processing (NLP). In addition, both global~\cite{dosovitskiy2020image} and local~\cite{liu2021swin} vision-Transformers have demonstrated impressive performance in the field of CV. This characteristic has inspired researchers to design better convolutional neural networks (CNNs) by utilizing larger convolution kernels. For example, ConvNeXt~\cite{liu2022convnet} uses large convolution kernels to obtain a larger receptive field and achieve comparable performance to Swin-Transformer. RepLKNet~\cite{ding2022scaling} scales up kernels to $31\times31$ using depth-wise convolution and re-parameterization, achieving comparable or superior results to Swin-Transformer on various tasks. VAN~\cite{guo2022visual} explores the effective application of attention mechanisms in CV and proposes a new large kernel attention (LKA) module. SLaK~\cite{liu2022more} proposes a recipe for applying extremely large kernels from the perspective of sparsity, allowing for the smooth scaling up of kernels to $61\times61$ with improved performance. Drawing inspiration from such designs, we developed a large kernel distillation block (LKDB) with large kernel attention (LKA) to further improve the representation ability of LKDN.

\subsection{Re-parameterization}
Re-parameterization is a piratical technique for designing lightweight models that can improve performance without increasing the inference burden. ACNet~\cite{ding2019acnet} employs asymmetric convolution to strengthen kernel structures, yielding better results than normal convolution. RepVGG~\cite{ding2021repvgg} decomposes a standard $3\times3$ convolution into a multi-branch topology comprising identity mapping, $1\times1$ convolution, and $3\times3$ convolution, enabling traditional VGG-style CNNs to achieve similar performance and faster inference speed than SOTA on several high-level vision tasks. Diverse Branch Block~\cite{ding2021diverse} combines diverse branches of varying scales and complexities to enrich the feature space, constructing a convolutional network unit resembling Inception~\cite{szegedy2015going}. MobileOne~\cite{vasu2022improved} leverages the re-parameterization technique to enhance the model's accuracy and speed, eventually achieving SOTA performance within effective architectures while being many times faster on mobile devices. As a result. we employed re-parameterization techniques to optimize the feature extraction block in LKDN.
\subsection{Adan Optimizer}

The Adam optimizer~\cite{kingma2014adam} is widely utilized in various deep learning domains. By utilizing first and second order gradient moment estimation, it dynamically adjusts the learning rate of each parameter to achieve faster convergence than stochastic gradient descent (SGD). However, Adam can also suffer from non-convergence~\cite{reddi2019convergence} and local optima~\cite{wilson2017marginal,keskar2017improving}. Recently, Adan optimizer~\cite{xie2022adan} combines modified Nesterov impulse, adaptive optimization, and decoupling weight attenuation. Adan uses extrapolation points to perceive gradient information beforehand, allowing for efficient escape from sharp local minima and increasing model generalization. Based on extensive experiments, it has been shown that the Adan optimizer outperforms existing SOTA optimizers for both CNNs and Transformers. Therefore, we aim to apply the Adan algorithm to lightweight super-resolution tasks.
\section{Method}
\subsection{Network Architecture}
Our approach follows the same framework design as BSRN~\cite{li2022blueprint}, as depicted in Figure~\ref{lkdn}. It comprises four components: shallow feature extraction, multiple stacked feature distillation blocks, multi-layer feature fusion, and image reconstruction block. 

Compared to traditional super-resolution models, our approach involves replicating the input image $n$ times during the pre-processing stage, followed by a concatenation of the replicated images. Given the input $I_{LR}$, this procedure can be expressed as:
\begin{equation}
	I^n_{LR}={\rm Concat}(I_{LR}^n),
\end{equation} where ${\rm Concat}(\cdot)$ denotes the concatenation operation along the channel dimension, and $n$ is the number of replicated input image $I_{LR}$. Then the initial feature extraction is implemented by a $3 \times 3$ BSConv to generate shallow features from the input LR image as:
$F_0=h_{ext}(I^n_{LR})$, where $h_{ext}(\cdot)$ denotes the module of shallow feature extraction, and $F_0$ denotes shallow feature. The structure of BSconv is shown in Figure~\ref{bsconv}, which consists of a $1\times1$ convolution and a depth-wise convolution.

The next part of LKDN is to extract deep features through a stack of LKDBs, which can be formulated as:
\begin{equation}
	F_k=H^m_{LKDB}(…H^1_{LKDB}(F_0)),1\leq k\leq m,
\end{equation}
where $H^k_{LKDB}(\cdot)$ denotes the $k$th LKDB, $m$ is the number of used LKDBs, and $F_{k}$ represents the output feature of the $k$th LKDB. 

After gradually refining by the LKDBs, all the intermediate features are fused and activated by a $1\times1$ convolution layer and a GELU~\cite{hendrycks2016gaussian} activation. A $3\times3$ BSConv layer is used to smooth the fused features. The process of multi-layer feature fusion can be formulated as:
\begin{equation}
	F_{fusion}=H_{fusion}({\rm Concat}(F_1,…,F_{k})),
\end{equation}
where $H_{fusion}(\cdot)$ denotes the feature fusion module, and $F_{fusion}$ is the fused features.

Finally, a skip connection is employed in the model to enhance the residual learning and the SR images are obtained through image reconstruction as:
\begin{equation}
	I_{SR}=H_{rec}(F_{fusion}+F_0),
\end{equation}
where $H_{rec}(\cdot)$ denotes the image reconstruction module, and $I_{SR}$ is the output of the model. The reconstruction process only includes a $3\times3$ convolution and pixel-shuffle operation~\cite{shi2016real}.

\subsection{Rethinking the BSRN}
The performance of BSRN models has been improved by ESA, CCA, and multiple residual connections. However, the complex structure results in lower computational efficiency. RLFN used a pruning sensitivity analysis tool based on a one-shot structured pruning algorithm to analyze the redundancy of the ESA block and discovered a significant amount of redundancy. Therefore, we removed the ESA and CCA modules of BSRN and introduced more efficient attention modules.

Recent studies~\cite{guo2022visual,zhou2023efficient} have shown that the performance of a model can be improved while maintaining acceptable complexity by using large kernel convolution reasonably. VAN~\cite{guo2022visual} utilizes convolution decomposition to split a large kernel convolution into three parts: a spatial local convolution, a spatial long-range convolution, and a channel convolution. Specifically, a $K\times K$ convolution is decomposed into a $(2d-1)\times(2d-1)$ depth-wise convolution, a $[\frac{K}{d}]\times[\frac{K}{d}]$ depth-wise dilation convolution with dilation $d$, and a $1\times1$ convolution. By decomposing large kernel convolution, the model can capture long-range relationships with minimal computational cost and parameters. Similar to VAPSR, we perform convolution decomposition in a different order, and as a result, the LKA module is shown in Figure~\ref{lka} and can be expressed as follows:
\begin{equation}
	X_{atten}={{\rm Conv}_{DW-D}({\rm Conv}_{DW}({\rm Conv}_{1\times1}}(F)),
\end{equation}
\begin{equation}
	Output=X_{atten}\otimes F,
\end{equation}
where ${\rm Conv}_{DW-D}(\cdot)$ and ${\rm Conv}_{DW}(\cdot)$ denotes dilated depth-wise convolution and depth-wise convolution respectively,  $X_{atten}$ denotes attention map, $\otimes$ denotes element-wise product operation, and $F$ denotes the input feature. By decomposing a large $17\times 17$ convolution into a $1\times1$ point-wise convolution, a depth-wise $5\times5$ convolution, and a depth-wise dilation convolution with a kernel size of $5$ and dilation of $3$, our model can reduce complexity and improve performance. Replacing ESA and CCA modules with LKA modules can also further improve inference speed.

\subsubsection{Re-parameterization}

According to~\cite{zhang2021edge}, excessive skip connection operations can increase memory access cost and inference time. However, the efficient separable distillation block (ESDB) used in BSRN includes four skip connections. To address this issue, we replace the residual connection of the ESDB with a re-parameterizable skip connection and add the branch of BSConv. To replace the Blueprint Shallow Residual Block (BSRB) in BSRN, we introduce the Re-parameterized Blueprint Shallow Block (RBSB) structure, as shown in Figure~\ref{rbsb}. Specifically, we introduce a re-parameterizable skip connection on the $1\times1$ point-wise convolution and the $3\times3$ depth-wise convolution. Additionally, we parallel a $1\times1$ depth-wise convolution on the $3\times3$ depth-wise convolution.

The re-parameterization process consists of two steps. First, the input feature $F_0$ is passed through a re-parameterized $1\times1$ point-wise convolution:
\begin{equation}
	F_1={\rm Conv}_{1\times1}(F_0)+F_0.
\end{equation}
Then, a re-parameterized $3\times3$ depth-wise convolution is applied:
\begin{equation}
	F_2={\rm Conv}_{DW_{3\times3}}(F_1)+{\rm Conv}_{D_{1\times1}}(F_1)+F_1.
\end{equation}
The expressions for $F_1$ and $F_2$ during inference using the re-parameterization technique are thus as follows:
\begin{align}
	\begin{split}
		F_1&={\rm Conv}_{1\times1}(F_0),\\
		F_2&={\rm Conv}_{DW_{3\times3}}(F_0).
	\end{split}
\end{align}

\subsubsection{Large Kernel Distillation Block}
After analyzing the characteristics of BSRN and VAPSR, we developed an even more efficient large kernel distillation block (LKDB). The complete structure of LKDB can be seen in Figure~\ref{lkdb}. It comprises four components: feature distillation, feature fusion, feature enhancement, and feature transformation. In the first stages, given the input $F_{in}$, the feature distillation operation can be described as:
\begin{align}
	\begin{split}
		F_{d_1},F_{r_1}&=D_1(F_{in}),R_1(F_{in}),\\
		F_{d_2},F_{r_2}&=D_2(F_{r_1}),R_2(F_{r_1}),\\
		F_{d_3},F_{r_3}&=D_3(F_{r_2}),R_3(F_{r_2}),\\
		F_{d_4}&=D_4(F_{r_3}),
	\end{split}
\end{align}
where $D_i$, $R_i$ denote the $i$th distillation and $i$th refinement layer, respectively. $F_{d_i}$, $F_{r_i}$ represents the $i$th distilled features and $i$th refined features, respectively. In the feature fusion stage, all the distilled features produced by previous distillation steps are concatenated together and then fused by a $1\times 1$ convolution as:
\begin{equation}
	F_{fusion}=H_{fusion}({\rm Concat}(F_{d_1},F_{d_2},F_{d_3},F_{d_4})),
\end{equation}
where $H_{fusion}$ denotes the $1\times1$ convolution layer, and $F_{fusion}$ is the fused feature. For the feature enhancement stage, we introduce an efficiency large kernel attention (LKA) block as:
\begin{equation}
	F_{enhance}=H_{LKA}(F_{fusion}),
\end{equation}
where $H_{LKA}$ denotes the LKA module, and $F_{enhance}$ is the enhanced feature. To enhance the performance of the model, we employ a $1\times 1$ convolution in the feature transformation stage, while a pixel normalization~\cite{zhou2023efficient} module is incorporated to ensure stable model training as:
\begin{equation}
	F_{trans}={\rm Norm_{pixel}}(H_{trans}(F_{enhanced})),
\end{equation}
where $H_{trans}$ denotes the $1\times1$ convolution transformation, $F_{trans}$ is the transformed feature, and ${\rm Norm_{pixel}}$ refers to the pixel normalization operation~\cite{zhou2023efficient}. Finally, a long skip connection is used to strengthen the residual learning ability of the model as:
\begin{equation}
	F_{out}=F_{trans}+F_{in}.
\end{equation}
\section{Experiments}
\subsection{Datasets and Evaluation Metrics}
We utilized a training set of 800 images from DIV2K~\cite{Agustsson_2017_CVPR_Workshops} and 2650 images from Flickr2K~\cite{Lim2017Enhanced}. Our evaluation of the models is performed on commonly used benchmark datasets, including Set5~\cite{Bevilacqua2012Low}, Set14~\cite{zeyde2010single}, B100~\cite{martin2001database}, Urban100~\cite{huang2015single}, and Manga109~\cite{matsui2017sketch}. The training data was augmented with random horizontal flips and 90-degree rotations. The evaluation metrics used are the average peak-signal-to-noise ratio (PSNR) and the structural similarity~\cite{Wang2004Image} (SSIM) on the luminance (Y) channel.
\subsection{Implementation details of LKDN}
The proposed LKDN model is composed of 8 LKDBs with a distillation structure channel number and attention module channel number set to 56, and is trained with BSB to reduce training time. The mini-batch size and input patch size for each LR input are set to 64 and $48\times 48$, respectively. We train the model using the common $L_1$ loss function and the Adan optimizer~\cite{xie2022adan}, with $\beta_1 = 0.98$, $\beta_2 = 0.92$ and $\beta_3 = 0.99$. To stabilize training, we set the exponential moving average (EMA) to 0.999. The learning rate is set to a constant $5\times10^{-3}$ for the entire $1\times10^{6}$ training iterations.

We propose a smaller version of LKDN, called LKDN-S, for the NTIRE 2023 Efficient SR Challenge~\cite{li2023ntire_esr}. LKDN-S comprises 5 LKDBs and 42 channels, and is trained with RBSB. We also employ re-parameterization techniques in the up-sample layer of LKDN-S. The training process of LKDN-S involves two stages: an initial training stage and a fine-tuning stage. In the initial training stage, we randomly crop 128 mini-batch HR patches with a size of $256\times256$. We train LKDN-S using the common L1 loss function with a learning rate of $5\times10^{-3}$ and $9.5\times10^5$ iterations. In the fine-tuning stage, we set the patch size of HR images and batch size to $480\times480$ and 64, respectively. LKDN-S is fine-tuned using the $L_2$ loss function with a learning rate of $2\times10^{-5}$, and a total of $5\times10^4$ iterations. The EMA is set to 0.999 and Adan optimizer~\cite{xie2022adan}, with $\beta_1 = 0.98$, $\beta_2 = 0.92$ and $\beta_3 = 0.99$ is applied in both stages.

We implement all our models using PyTorch 1.11 and Nvidia GeForce RTX 3080 GPUs.
\begin{table*}
	\centering
	\caption{Ablation study on large kernel attention.}
	\vspace{-0.3cm}
	\label{abla_lka}
	\scalebox{0.8}{\hspace{-0.28cm}
		\begin{tabular}{|c|c|c|c|c|c|c|c|} 
			\hline
			Method                   & Params[K] & Multi-Adds[G] & Set5~\cite{Bevilacqua2012Low}        & Set14~\cite{zeyde2010single}         & B100~\cite{martin2001database}         & Urban100~\cite{huang2015single}     & Manga109~\cite{huang2015single}      \\ 
			\hline
			BSRN                     & 352       & 19.4          & 32.29 / 0.8950 & 28.63 / 0.7824 & 27.59 / 0.7365 & 26.12 / 0.7870 & 30.58/0.9091  \\ 
			\hline
			BSRN-w/o ESA\&CCA  & 315       & 18.2          & 32.13 / 0.8945 & 28.56 / 0.7814 & 27.54 / 0.7355 & 25.92 / 0.7811 & 30.31 / 0.9071  \\ 
			\hline
			LKDN\_\textit{C}64\_\textit{A}32      & 322       & 18.4          & \underline{32.28} / \textbf{0.8963} & \underline{28.69} / \underline{0.7835} & \underline{27.63} / \underline{0.7381} & \underline{26.19} / \underline{0.7894} & \underline{30.72} / \underline{0.9113}  \\ 
			\hline
			LKDN\_\textit{C}56\_\textit{A}56      & 322       & 18.3          & \textbf{32.30} / \underline{0.8962} & \textbf{28.70} / \textbf{0.7838} & \textbf{26.65} / \textbf{0.7385} & \textbf{26.22} / \textbf{0.7901} & \textbf{30.76} / \textbf{0.9114}  \\
			\hline
	\end{tabular}}

\end{table*}
\begin{table*}
	\centering
	\caption{PSNR / SSIM comparison of different basic blocks in the feature distillation connections of LKDN-S.}
	\vspace{-0.3cm}
	\label{rep_table}
	\scalebox{0.8}{
		\begin{tabular}{|c|c|c|c|c|c|} 
			\hline
			Method & Set5~\cite{Bevilacqua2012Low}        & Set14~\cite{zeyde2010single}         & B100~\cite{martin2001database}         & Urban100~\cite{huang2015single}     & Manga109~\cite{huang2015single}      \\ 
			\hline
			BSRB   & 32.05 / 0.8932 & 28.52 / 0.7801 & 27.53 / 0.7349 & 25.85 / 0.7787 & 30.23 / 0.9050  \\ 
			\hline
			BSB    & 32.06 / 0.8935 & 28.54 / \textbf{0.7805} & 27.54 / 0.7352 & 25.89 / 0.7797 & \textbf{30.29} / 0.9057  \\ 
			\hline
			RBSB   & \textbf{32.11} / \textbf{0.8936} & \textbf{28.55} / \textbf{0.7805} & \textbf{27.55} / \textbf{0.7354} & \textbf{25.90} / \textbf{0.7803} & \textbf{30.29} / \textbf{0.9060}  \\
			\hline
	\end{tabular}}
	
\end{table*}

\begin{table*}[h!]
	\caption{PSNR / SSIM comparison of applying Adam~\cite{kingma2014adam} and Adan~\cite{xie2022adan} optimizers.}

	\centering
	\scalebox{0.8}{
		\begin{tabular}{|c|c|c|c|c|c|c|} 
			\hline
			Method  &  Training-Time[h]  &Set5~\cite{Bevilacqua2012Low}        & Set14~\cite{zeyde2010single}         & B100~\cite{martin2001database}         & Urban100~\cite{huang2015single}     & Manga109~\cite{huang2015single}   \\ 
			\hline
			LKDN\_Adam          & 50                              & \textbf{32.41} / 0.8975          & 28.77 / 0.7854              & \textbf{27.69} / 0.7399    & 26.36 / 0.7949                & 30.93 / 0.9132        \\ 
			\hline
			LKDN\_Adan         & 45 $\downarrow$                              & 32.39 / \textbf{0.8979}             & \textbf{28.79} / \textbf{0.7859}              & \textbf{27.69} / \textbf{0.7402}    & \textbf{26.42} / \textbf{0.7965}                & \textbf{30.97} / \textbf{0.9140}        \\
			\hline
	\end{tabular}}
\label{optm}

\end{table*}
\begin{table}
	\centering
	\caption{Following~\cite{li2022ntire}, we compare the computational costs.}
	\vspace{-0.3cm}
	\setlength\tabcolsep{4pt}
	\scalebox{0.74}{
		\begin{tabular}{|c|c|c|c|c|} 
			\hline
			Method & DIV2K Val[dB] & Params[K] & Multi-Adds[G] & Runtime[ms]  \\ 
			\hline
			BSRN~\cite{li2022blueprint}   & 29.07          & 352       & 19.4          & 83.75        \\ 
			\hline
			VAPSR~\cite{zhou2023efficient}  & 29.15          & 342       & 19.5          & 105.82       \\ 
			\hline
			LKDN   & 29.18        & 322       & 18.3         & 85.65      \\
			\hline
	\end{tabular}}
\label{computa}
	
\end{table}

\subsection{Ablation Study}

\begin{table*}
	\caption{Quantitative comparison with state-of-the-art methods on benchmark datasets. ’Multi-Adds’ is calculated with a $1280\times720$ GT image. The best and second best are in \textcolor{red}{red} and \textcolor{blue}{blue} respectively.}
	\centering\label{benchmark}
	\setlength\tabcolsep{3pt}
	\scalebox{0.92}{
		\begin{tabular}{|c|c|c|c|c|c|c|c|c|} 
			\hline
			Method  & Scale & Params.[K] & Multi-Adds[G] & Set5~\cite{Bevilacqua2012Low}         & Set14~\cite{zeyde2010single}        & B100~\cite{martin2001database}           & Urban100~\cite{huang2015single}       & Manga109~\cite{matsui2017sketch}       
			\\[0.5ex] \hline\hline
			Bicubic &    $\times2$   & -      & -          & 33.66 / 0.9299 & 30.24 / 0.8688 & 29.56 / 0.8431   & 26.88 / 0.8403   & 30.80 / 0.9339  \\
			SRCNN~\cite{Dong2015Image}   &  $\times2$     & 8      & 52.7       & 36.66 / 0.9542 & 32.45 / 0.9067 & 31.36 / 0.8879   & 29.50 / 0.8946   & 35.60 / 0.9663  \\
			FSRCNN~\cite{dong2016accelerating}  &    $\times2$   & 13     & 6.0        & 37.00 / 0.9558 & 32.63 / 0.9088 & 31.53 / 0.8920   & 29.88 / 0.9020   & 36.67 / 0.9710  \\
			VDSR~\cite{Kim2016Accurate}    &    $\times2$   & 666    & 612.6      & 37.53 / 0.9587 & 33.03 / 0.9124 & 31.90 / 0.8960   & 30.76 / 0.9140   & 37.22 / 0.9750  \\
			LapSRN~\cite{lai2017deep}  &   $\times2$    & 251    & 29.9       & 37.52 / 0.9591 & 32.99 / 0.9124 & 31.80 / 0.8952   & 30.41 / 0.9103   & 37.27 / 0.9740  \\ 
			DRRN~\cite{Tai2017Image}    &   $\times2$    & 298    & 6796.9     & 37.74 / 0.9591 & 33.23 / 0.9136 & 32.05 / 0.8973   & 31.23 / 0.9188   & 37.88 / 0.9749  \\ 
			MemNet~\cite{tai2017memnet}  &    $\times2$   & 678    & 2662.4     & 37.78 / 0.9597 & 33.28 / 0.9142 & 32.08 / 0.8978 & 31.31 / 0.9195   & 37.72 / 0.9740  \\
			IDN~\cite{hui2018fast}     &    $\times2$   & 553    & 124.6      & 37.83 / 0.9600 & 33.30 / 0.9148 & 32.08 / 0.8985   & 31.27 / 0.9196   & 38.01 / 0.9749  \\
			CARN~\cite{ahn2018fast}    &    $\times2$   & 1592   & 222.8      & 37.76 / 0.9590 & 33.52 / 0.9166 & 32.09 / 0.8978   & 31.92 / 0.9256   & 38.36 / 0.9765  \\
			IMDN~\cite{hui2019lightweight}    &   $\times2$    & 694    & 158.8      & 38.00 / 0.9605 & 33.63 / 0.9177 & 32.19 / 0.8996   & 32.17 / 0.9283   & 38.88 / 0.9774  \\
			PAN~\cite{zhao2020efficient}     &   $\times2$    & 261    & 70.5       & 38.00 / 0.9605 & 33.59 / 0.9181 & 32.18 / 0.8997   & 32.01 / 0.9273   & 38.70 / 0.9773  \\
			LAPAR-A~\cite{li2020lapar} &  $\times2$     & 548    & 171.0      & 38.01 / 0.9605 & 33.62 / 0.9183 & 32.19 / 0.8999   & 32.10 / 0.9283   & 38.67 / 0.9772  \\
			RFDN~\cite{liu2020residual}    &   $\times2$    & 535    & 95.0       & 38.05 / 0.9606 & 33.68 / 0.9184 & 32.16 / 0.8994   & 32.12 / 0.9278   & 38.88 / 0.9773  \\
			RFLN~\cite{Kong2022Residual}    &  $\times2$     & 527    & 115.4      & 38.07 / 0.9607 & 33.72 / 0.9187 & 32.22 / 0.9000   & 32.33 / 0.9299   & -             \\
			BSRN~\cite{li2022blueprint}    &    $\times2$   & 332    & 73.0       & \textcolor{blue}{38.10} / 0.9610 & 33.74 / 0.9193 & 32.24 / 0.9006   & 32.34 / 0.9303   & \textcolor{blue}{39.14} / \textcolor{blue}{0.9782}  \\
			VAPSR~\cite{zhou2023efficient}   &   $\times2$    & 329    & 74.0       & 38.08 / \textcolor{red}{0.9612} & \textcolor{blue}{33.77} / \textcolor{blue}{0.9195} & \textcolor{red}{32.27} / \textcolor{red}{0.9011}   & \textcolor{blue}{32.45} / \textcolor{blue}{0.9316}   & -             \\
			LKDN    &  $\times2$     &   304     &    69.1        &     \textcolor{red}{38.12} / \textcolor{blue}{0.9611}         &      \textcolor{red}{33.90} / \textcolor{red}{0.9202}	        &      \textcolor{red}{32.27} / \textcolor{blue}{0.9010}         &       \textcolor{red}{32.53} / \textcolor{red}{0.9322}	         &   \textcolor{red}{39.19} / \textcolor{red}{0.9784}            \\[0.5ex] \hline\hline
			Bicubic &   $\times3$    & -      & -          & 30.39 / 0.8682 & 27.55 / 0.7742 & 27.21 / 0.7385 & 24.46 / 0.7349 & 26.95 / 0.8556  \\
			SRCNN~\cite{Dong2015Image}   &    $\times3$   & 8      & 52.7       & 32.75 / 0.9090 & 29.30 / 0.8215 & 28.41 / 0.7863   & 26.24 / 0.7989   & 30.48 / 0.9117  \\
			FSRCNN~\cite{dong2016accelerating}  &   $\times3$    & 13     & 5.0        & 33.18 / 0.9140 & 29.37 / 0.8240 & 28.53 / 0.7910   & 26.43 / 0.8080   & 31.10 / 0.9210  \\
			VDSR~\cite{Kim2016Accurate}    &    $\times3$   & 666    & 612.6      & 33.66 / 0.9213 & 29.77 / 0.8314 & 28.82 / 0.7976   & 27.14 / 0.8279   & 32.01 / 0.9340  \\
			DRRN~\cite{Tai2017Image}    &   $\times3$    & 298    & 6796.9     & 34.03 / 0.9244 & 29.96 / 0.8349 & 28.95 / 0.8004   & 27.53 / 0.8378   & 32.71 / 0.9379  \\
			MemNet~\cite{tai2017memnet}  &   $\times3$    & 678    & 2662.4     & 34.09 / 0.9248 & 30.00 / 0.8350 & 28.96 / 0.8001   & 27.56 / 0.8376 & 32.51 / 0.9369  \\
			IDN~\cite{hui2018fast}     &    $\times3$   & 553    & 56.3       & 34.11 / 0.9253 & 29.99 / 0.8354 & 28.95 / 0.8013   & 27.42 / 0.8359   & 32.71 / 0.9381  \\
			CARN~\cite{ahn2018fast}    &   $\times3$    & 1592   & 118.8      & 34.29 / 0.9255 & 30.29 / 0.8407 & 29.06 / 0.8034   & 28.06 / 0.8493   & 33.50 / 0.9440  \\
			IMDN~\cite{hui2019lightweight}    &  $\times3$     & 703    & 71.5       & 34.36 / 0.9270 & 30.32 / 0.8417 & 29.09 / 0.8046   & 28.17 / 0.8519   & 33.61 / 0.9445  \\
			PAN~\cite{zhao2020efficient}     &  $\times3$     & 261    & 39.0       & 34.40 / 0.9271 & 30.36 / 0.8423 & 29.11 / 0.8050   & 28.11 / 0.8511   & 33.61 / 0.9448  \\
			LAPAR-A~\cite{li2020lapar} &    $\times3$   & 544    & 114.0      & 34.36 / 0.9267 & 30.34 / 0.8421 & 29.11 / 0.8054   & 28.15 / 0.8523   & 33.51 / 0.9441  \\
			RFDN~\cite{liu2020residual}    &  $\times3$     & 541    & 42.2       & 34.41 / 0.9273 & 30.34 / 0.8420 & 29.09 / 0.8050   & 28.21 / 0.8525   & 33.67 / 0.9449  \\
			BSRN~\cite{li2022blueprint}    &   $\times3$    & 340    & 33.3       & 34.46 / 0.9277 & 30.47 / 0.8449 & 29.18 / 0.8068   & 28.39 / 0.8567   & \textcolor{blue}{34.05} / \textcolor{blue}{0.9471}  \\
			VAPSR~\cite{zhou2023efficient}   &   $\times3$    & 337    & 33.6       & \textcolor{blue}{34.52} / \textcolor{blue}{0.9284} & \textcolor{red}{30.53} / \textcolor{blue}{0.8452} & \textcolor{blue}{29.19} / \textcolor{blue}{0.8077}   & \textcolor{blue}{28.43} / \textcolor{blue}{0.8583}   & -             \\
			LKDN    &   $\times3$    &   311     &      31.4      &     \textcolor{red}{34.54} / \textcolor{red}{0.9285}         &      \textcolor{blue}{30.52} / \textcolor{red}{0.8455}        &       \textcolor{red}{29.21} / \textcolor{red}{0.8078}	         &         \textcolor{red}{28.50} / \textcolor{red}{0.8601}       &    \textcolor{red}{34.08} / \textcolor{red}{0.9475}           \\[0.5ex] \hline\hline
			Bicubic &    $\times4$   & -      & -          & 28.42 / 0.8104 & 26.00 / 0.7027 & 25.96 / 0.6675   & 23.14 / 0.6577   & 24.89 / 0.7866  \\
			SRCNN~\cite{Dong2015Image}   &   $\times4$     & 8      & 52.7       & 30.48 / 0.8626 & 27.50 / 0.7513 & 26.90 / 0.7101   & 24.52 / 0.7221   & 27.58 / 0.8555  \\
			FSRCNN~\cite{dong2016accelerating}  &   $\times4$     & 13     & 4.6        & 30.72 / 0.8660 & 27.61 / 0.7550 & 26.98 / 0.7150   & 24.62 / 0.7280   & 27.90 / 0.8610  \\
			VDSR~\cite{Kim2016Accurate}    &   $\times4$     & 666    & 612.6      & 31.35 / 0.8838 & 28.01 / 0.7674 & 27.29 / 0.7251   & 25.18 / 0.7524   & 28.83 / 0.8870  \\
			LapSRN~\cite{lai2017deep}  &    $\times4$    & 813    & 149.4      & 31.54 / 0.8852 & 28.09 / 0.7700 & 27.32 / 0.7275   & 25.21 / 0.7562 & 29.09 / 0.8900  \\
			DRRN~\cite{Tai2017Image}    &   $\times4$     & 298    & 6796.9     & 31.68 / 0.8888 & 28.21 / 0.7720 & 27.38 / 0.7284   & 25.44 / 0.7638   & 29.45 / 0.8946  \\
			MemNet~\cite{tai2017memnet}  &   $\times4$     & 678    & 2662.4     & 31.74 / 0.8893 & 28.26 / 0.7723 & 27.40 / 0.7281   & 25.50 / 0.7630   & 29.42 / 0.8942  \\
			IDN~\cite{hui2018fast}     &   $\times4$     & 553    & 32.3       & 31.82 / 0.8903 & 28.25 / 0.7730 & 27.41 / 0.7297   & 25.41 / 0.7632   & 29.41 / 0.8942  \\
			CARN~\cite{ahn2018fast}     &   $\times4$     & 1592   & 90.9       & 32.13 / 0.8937 & 28.60 / 0.7806 & 27.58 / 0.7349   & 26.07 / 0.7837   & 30.47 / 0.9084  \\
			IMDN~\cite{hui2019lightweight}    &  $\times4$      & 715    & 40.9       & 32.21 / 0.8948 & 28.58 / 0.7811 & 27.56 / 0.7353   & 26.04 / 0.7838   & 30.45 / 0.9075  \\
			PAN~\cite{zhao2020efficient}     &   $\times4$     & 272    & 28.2       & 32.13 / 0.8948 & 28.61 / 0.7822 & 27.59 / 0.7363   & 26.11 / 0.7854   & 30.51 / 0.9095  \\
			LAPAR-A~\cite{li2020lapar} &   $\times4$     & 659    & 94.0       & 32.15 / 0.8944 & 28.61 / 0.7818 & 27.61 / 0.7366   & 26.14 / 0.7871   & 30.42 / 0.9074  \\
			RFDN~\cite{liu2020residual}    &    $\times4$    & 550    & 23.9       & 32.24 / 0.8952 & 28.61 / 0.7819 & 27.57 / 0.7360   & 26.11 / 0.7858 & 30.58 / 0.9089  \\
			RLFN~\cite{Kong2022Residual}    &   $\times4$     & 543    & 29.8       & 32.24 / 0.8952 & 28.62 / 0.7813 & 27.60 / 0.7364   & 26.17 / 0.7877   & -             \\
			BSRN~\cite{li2022blueprint}    &   $\times4$     & 352    & 19.4       & 32.35 / 0.8966 & 28.73 / 0.7847 & 27.65 / 0.7387   & 26.27 / 0.7908   & 30.84 / 0.9123  \\
			VAPSR~\cite{zhou2023efficient}   &   $\times4$     & 342    & 19.5       & \textcolor{blue}{32.38} / \textcolor{blue}{0.8978} & \textcolor{blue}{28.77} / \textcolor{blue}{0.7852} & \textcolor{blue}{27.68} / \textcolor{blue}{0.7398}   & \textcolor{blue}{26.35} / \textcolor{blue}{0.7941}   & \textcolor{blue}{30.89} / \textcolor{blue}{0.9132}            \\
			LKDN-S    &  $\times4$      &   129     &    7.3       &      32.10 / 0.8938        &    28.62 / 0.7821          &      27.59 / 0.7371          &      26.07 / 0.7845          &       30.50 / 0.9078     \\
			LKDN    &  $\times4$      &   322     &    18.3        &      \textcolor{red}{32.39} / \textcolor{red}{0.8979}        &    \textcolor{red}{28.79} / \textcolor{red}{0.7859}          &      \textcolor{red}{27.69} / \textcolor{red}{0.7402}          &      \textcolor{red}{26.42} / \textcolor{red}{0.7965}          &       \textcolor{red}{30.97} / \textcolor{red}{0.9140}        \\
			\hline
	\end{tabular}}

\end{table*}

\subsubsection{Large Kernel Attention}
We conducted ablation studies to verify the efficacy of the LKA module, as presented in Table~\ref{abla_lka}. In this table, $C$ and $A$ denote the input channels of the distillation structure and attention module, respectively. Removing the ESA and CCA modules from BSRN resulted in a significant drop in model performance. However, by utilizing the LKA module, the network's receptive field increased, leading to an improvement in model performance while maintaining lower parameter and computation costs than the original BSRN model. Further performance improvements were achieved by adjusting the channel numbers of the distillation structure and attention module. Compared to the original BSRN, LKDN can achieve performance gains of more than \textbf{0.1 dB} on the Urban100~\cite{huang2015single} and Manga109~\cite{huang2015single} while maintaining lower parameter and computation costs.

\subsubsection{Re-parameterization}
\begin{figure}[!htbp]	
	\vspace{-1cm}
	\setlength{\abovecaptionskip}{3mm}
	\begin{subfigure}{1\linewidth}
		\vspace{1cm}
		\centering
		\includegraphics[scale=0.8]{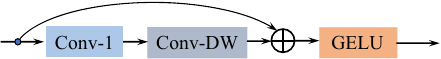}
		\caption{BSRB}
		\label{bsrb}
	\end{subfigure}

	\begin{subfigure}{1\linewidth}
	\vspace{0.8cm}
	\centering
	\includegraphics[scale=0.8]{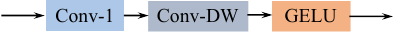}
	\caption{BSB}
	\label{bsb}
\end{subfigure}
	\caption{Non-reparameteried components. (a) BSRB: Blueprint Shallow Residual Block; (b) BSB: Blueprint Shallow Block.}
\end{figure}

To demonstrate the effectiveness of the proposed RBSB in Figure~\ref{rbsb}, we replaced the basic blocks in the feature distillation structure for comparison. Figure~\ref{bsrb} displays the BSRB used in BSRN, while Figure~\ref{bsb} illustrates the blueprint shallow block (BSB) obtained by directly removing the residual connections. The evaluation results we conducted on LKDN-S are presented in Table~\ref{rep_table}, demonstrating that eliminating unnecessary residual connections can improve performance while reducing model complexity. Re-parameterization can thus further improve performance while maintaining model complexity.

\begin{figure}[!htbp]
	\setlength{\abovecaptionskip}{3mm}
	
	\centering
	\includegraphics[scale=0.4]{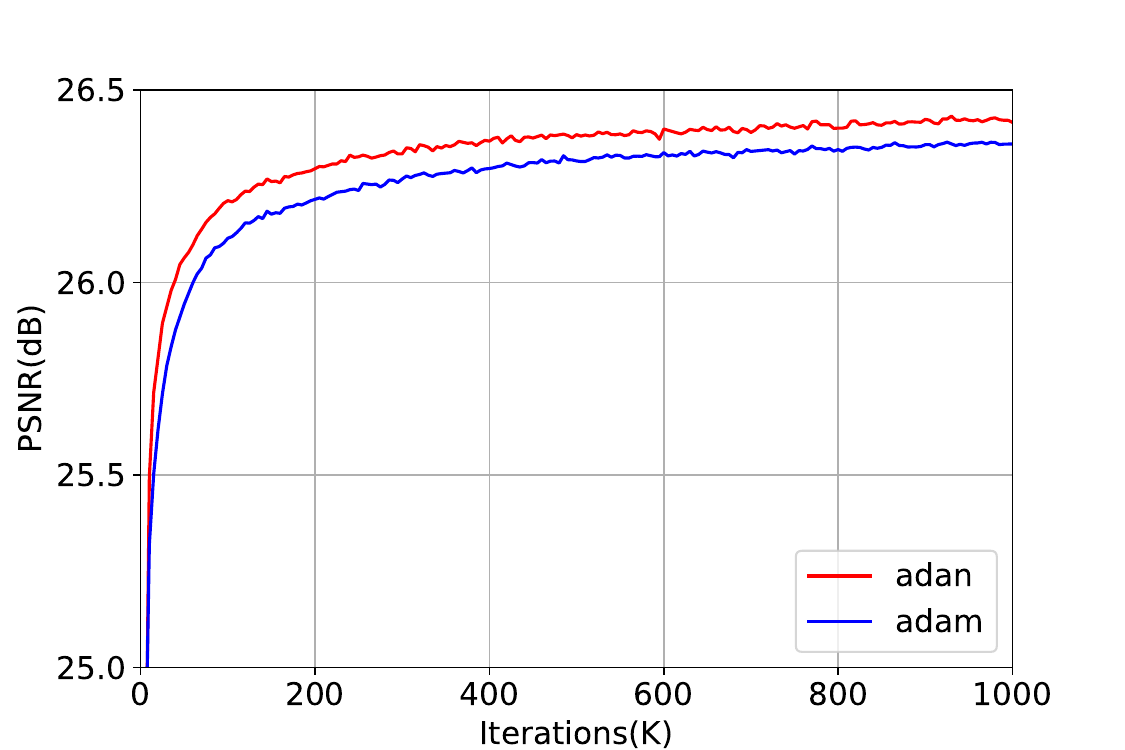}
	\caption{Convergence comparison between Adan~\cite{xie2022adan} and Adam~\cite{kingma2014adam} optimizers, using Ubran100~\cite{huang2015single} SR($\times4$).}
	\label{conver}
	
\end{figure}
\subsubsection{Optimizer}
The previous optimization of SR models primarily relied on the Adam optimizer~\cite{kingma2014adam}. However, the Adan optimizer~\cite{xie2022adan}, which has recently achieved state-of-the-art results on various vision tasks, has piqued the interest of researchers in the field. Therefore, we investigated the effects of the Adan optimizer on SR tasks. The results in Table~\ref{optm} show that using the Adan optimizer yields a training speedup of roughly \textbf{10\%} compared to using the Adam optimizer, with a significant performance improvement on various benchmark datasets. Moreover, as demonstrated in Figure~\ref{conver}, the Adan optimizer leads to faster convergence.

\subsection{Comparison with the State-of-the-art Methods}
\begin{figure*}[!htbp]
	\centering
	\includegraphics[scale=1.15]{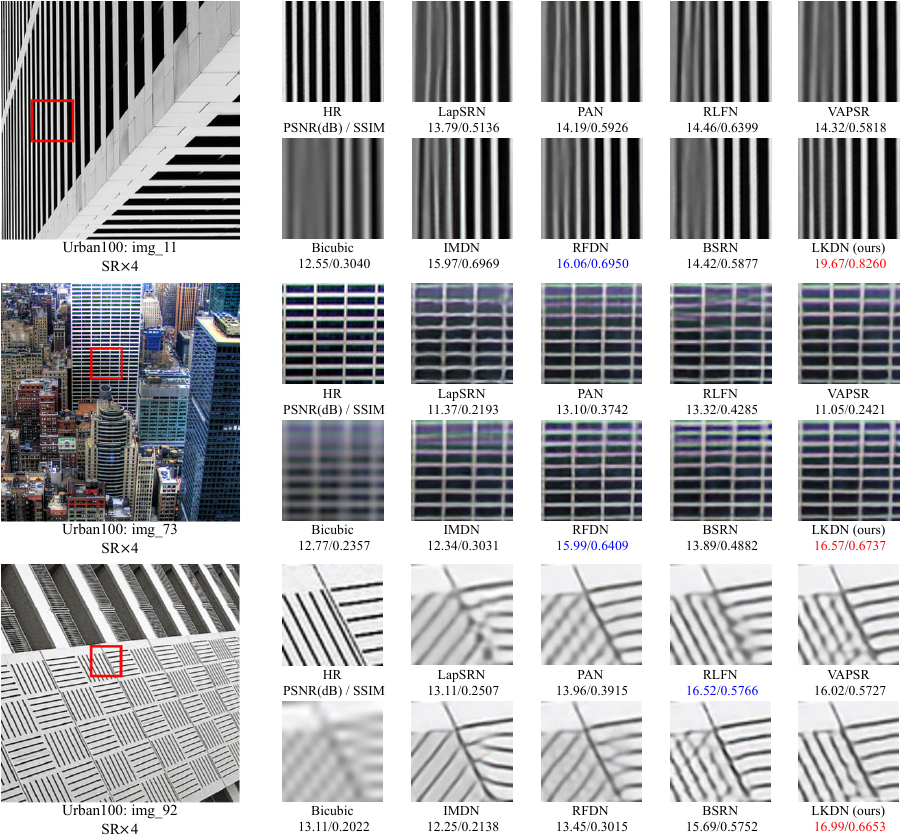}

	\caption{Qualitative and quantitative comparison on SR ($\times4$), the best and second best are in \textcolor{red}{red} and \textcolor{blue}{blue} respectively.}
	\label{vis}
	
\end{figure*}
We have evaluated our proposed LKDN with various state-of-the-art lightweight SR methods, which includes FSRCNN~\cite{dong2016accelerating}, VDSR~\cite{Kim2016Accurate}, LapSRN~\cite{lai2017deep}, DRRN~\cite{Tai2017Image}, MemNet~\cite{tai2017memnet}, IDN~\cite{hui2018fast}, CARN~\cite{ahn2018fast}, IMDN~\cite{hui2019lightweight}, PAN~\cite{zhao2020efficient}, LAPAR-A~\cite{li2020lapar}, RFDN~\cite{liu2020residual}, RFLN~\cite{Kong2022Residual}, BSRN~\cite{li2022blueprint}, and VAPSR~\cite{zhou2023efficient}. These methods have been compared for upscale factors of $\times2$, $\times3$, and $\times4$. Table~\ref{benchmark} presents the quantitative comparison results for these methods. With the incorporation of efficient attention modules, LKDN has outperformed other methods in terms of achieving the best performance while maintaining a lightweight model. In addition, our proposed solution for NTIRE 2023 Challenge, LKDN-S, has achieved competitive performance with only $129$K parameters and $7.3$G Multi-Adds for SR $\times4$.

Figure~\ref{vis} presents a qualitative comparison of our proposed method. The results indicate that our model is more capable of reconstructing high-similarity structures than existing methods. As an example, in images "img\_11" and "img\_92", existing methods typically generate obvious distortion and blurriness, while our approach accurately captures the lines. Furthermore, in "img\_73", VAPSR calculates the incorrect number of windows, resulting in a significant decrease in PSNR and SSIM, while our method can accurately restore the number of windows.

Table~\ref{computa} provides a more in-depth analysis of BSRN, VAPSR, and LKDN. The results show that LKDN outperforms BSRN while maintaining a comparable inference speed. Moreover, LKDN achieves faster inference speeds than VAPSR while maintaining superior performance.

\section{Conclusion}
We propose the Large Kernel Distillation Network (LKDN) as an efficient single-image super-resolution (SISR) solution. Through careful analysis of some state-of-the-art lightweight models, we identified their weaknesses and improved upon them to enhance the performance of LKDN. By incorporating the large kernel design, simplifying the model structure, introducing more efficient attention modules, and employing re-parameterization, we achieve a balance between performance and computational efficiency. A new optimizer is also introduced to accelerate the convergence of LKDN. Extensive experiments demonstrate that LKDN outperforms state-of-the-art efficient SR methods in terms of performance, parameters, and Multi-Adds.

\section{Acknowledgment} This work is supported in part by National Natural Science Foundation of China (62102330) and in part by Natural Science Foundation of Sichuan Province (2022NSFSC0947, 2022NSFSC0945).

{\small
	\bibliographystyle{ieee_fullname}
	\bibliography{egbib}
}

\end{document}